\documentclass{elsart3p}

\def\RR{{\rm I\kern-.1567em R}}                              
\def\ZZ{{\sf Z\kern-4.5pt Z}}

\usepackage{graphics, graphicx}
\usepackage{seceqn}

\journal{Nuclear Physics B}

\begin{document}
\begin{frontmatter}

\title{Pullback of the Volume Form, Integrable Models in Higher
Dimensions and Exotic Textures}

\vspace*{1cm}

\author[san]{C. ~Adam}
\ead{adam@fpaxp1.usc.es}
\author[san]{P. ~Klimas }
\ead{klimas@fpaxp1.usc.es}
\author[san]{J. ~S\'{a}nchez-Guill\'{e}n}
\ead{joaquin@fpaxp1.usc.es}

\address[san]{Departamento de Fisica de Particulas, Universidad
       de Santiago and Instituto Galego de Fisica de Altas Enerxias
       (IGFAE) E-15782 Santiago de Compostela, Spain}

\author[were1,were2]{A. ~Wereszczy\'{n}ski}
\ead{wereszczynski@th.if.uj.edu.pl}

\address[were1]{The Nils Bohr Institute, Copenhagen University,
Blegdamsvej 17, DK-2100 Copenhagen {\O},
       Denmark}

\address[were2]{Institute of Physics,  Jagiellonian University,
       Reymonta 4, 30-059 Krak\'{o}w, Poland
}

\begin{abstract}
A procedure allowing for the construction of Lorentz invariant
integrable models living in $d+1$ dimensional space-time and with
an $n$ dimensional target space is provided. Here,
integrability is understood as the existence of the generalized
zero-curvature formulation and infinitely many conserved
quantities. A close relation between the Lagrange density of the
integrable models and the pullback of the pertinent volume form on
target space is established. Moreover, we show that the conserved
currents are Noether currents generated by the volume preserving
diffeomorphisms. Further, we show how such models may emerge via abelian
projection of some gauge theories.
\\
Then we apply this framework to the construction of integrable models with
exotic textures. Particularly, we consider integrable models
providing exact suspended Hopf maps i.e., solitons with a nontrivial
topological charge of $\pi_4(S^3)\cong \ZZ_2$.
\\
Finally,  some families of integrable models with solitons of
$\pi_n(S^n)$ type are constructed. Infinitely many exact solutions
with arbitrary value of the topological index are found. In addition, we
demonstrate that they
saturate a Bogomolny bound.
\end{abstract}


\end{frontmatter}

\onecolumn

\section{Introduction}
Integrability has proven a valuable concept for the analysis and
solution of nonlinear field theories in 1+1 dimensions, but its generalization
to higher dimensions is a rather difficult endeavour, and a generally accepted
concept of higher-dimensional integrability does not yet exist.
One possible way to generalize integrability to higher dimensions was
proposed in \cite{joaquin1}, where the zero curvature of Zakharov and Shabat
has been generalized to higher dimensions. Further, it was demonstrated in the
same paper that some known higher-dimensional nonlinear field theories
possess the generalized zero curvature representation and,
at the same time, infinitely many
conservation laws.
\\
It is the main purpose of the present
paper to further develop this investigation.
We shall explicitly construct different families of higher-dimensional
field theories which possess the generalized zero curvature representation.
Further, we will find their infinitely many conservation laws as well as their
exact soliton solutions. A key ingredient will be
that their Lagrangians are related
to the volume forms on their respective target spaces, and the conserved
currents are, in turn, related to the volume preserving diffeomorphisms.
Before introducing our investigation,
it will be useful to review some known results in
order to facilitate some background for the constructions that follow.
\\
Shortly after the general proposal of \cite{joaquin1},
some new nonlinear field
theories in 3+1 dimensions with
two dimensional target space and possessing the generalized zero curvature
representation were constructed explicitly.
One first model was
introduced by Aratyn, Ferreira and Zimerman
(AFZ), and they explicitly constructed both infinitely many conservation laws
and infinitely many soliton solutions  \cite{afz1},
\cite{afz2}. Due to the two-dimensional target space, their solitons are,
in fact, topological and are classified by the Hopf index.
Some integrable generalizations of the model of AFZ were discussed in
\cite{aw1}, where again infinitely many conservation laws and infinitely
many topological (Hopf) solitons were found.
Another model giving rise to Hopf solitons had been originally proposed by
Nicole, who found its simplest Hopf soliton with Hopf index one \cite{nicole}.
This Nicole model again possesses the generalized curvature representation
\cite{S-2-syms}, but it only gives rise to finitely many conservation laws.
Only a submodel of the Nicole model, defined by further first order
equations (``integrability conditions'') in addition to the Euler--Lagrange
equations, possesses infinitely many conservation laws. Further,
for the Nicole model only the simplest soliton may be found in an analytic
form. Higher solitons have to be calculated numerically \cite{we-nicole}.
Some generalization of the Nicole model have been discussed in
\cite{aw2}, \cite{aw3},
where again only one analytical soliton solution could be found
in each model. Both the AFZ model and the Nicole model (and their
generalizations) allow for static finite energy solutions because their
kinetic term is chosen non-polynomial in order to have a scale invariant
energy and avoid Derrick's theorem. This idea of non-polynomial Lagrangians
is originally due to \cite{deser}, where it was applied to a phenomenological
model of pions.
A more geometric understanding of the conservation laws in the models
mentioned above was developed in  \cite{diffeo1}, \cite{babelon},
\cite{S-2-syms}, \cite{we abel},
where it was shown that the conserved currents are
just the Noether currents of the area-preserving diffeomorphisms on
target space. The off-shell divergence of these currents is proportional
to the Euler--Lagrange equations (for AFZ type models) or to a linear
combination of Euler--Lagrange equations and integrability conditions
(for Nicole type models), respectively.
In a parallel development, the generalized zero curvature representation,
integrability and conservation laws of chiral and non-linear sigma models
were investigated, e.g., in \cite{joaquin2}, \cite{ferreira}, \cite{fujii},
\cite{suzuki}.
\\
Soon after this, the investigation was extended to field theories
with a three dimensional target space, the best-known of which is
the Skyrme model \cite{skyrme1}, \cite{skyrme2}. The Skyrme model
again possesses the generalized zero curvature representation, but
only a finite number of conserved currents. But, again, there
exists a submodel of the Skyrme model which has infinitely many
conservation laws \cite{joaquin S3}. A detailed classification of
the integrability of field theories with three dimensional target
spaces has been performed in \cite{ASGW1 S3}, \cite{ASGW2 S3}. The
abelian projection of SU(2) Yang--Mills dilaton theory, which
effectively has a three dimensional target space, was studied in
\cite{we YMdil}. Integrable theories with higher dimensional
target spaces were investigated, e.g., in  \cite{ferreira},
\cite{ASGW2 S3}. Further, the concept of generalized curvature
representations and integrability was applied to non-linear sigma
models on noncommutative space-time in \cite{Kurkcuoglu}.
\\
After this brief review we will give the outline of the present paper, which
combines and generalizes the ingredients described above: both an algebraic
and a geometrical formulation of the generalized integrability, as well as
the analysis of the geometry and topology of the target space and its
interrelation with the symmetries and conservation laws of the field theories
under investigation.
\\
Our paper is organized as follows. First we briefly recall the idea
of the generalized integrability and discuss it in the case of $S^2$
target space, concretely for the AFZ model. We find the geometric condition
which makes this model integrable. In section 3 we extend this
geometric approach to models with more complicated target space
manifolds. We show how one can construct integrable models based on the
volume form on target space. We find a
family of infinitely many conserved quantities and explain their
existence by relating them with the pertinent symmetries of the
target space. Section 4 is devoted to the generalization to even higher
dimensional target spaces.
In Section 5 we explain how the models presented in the previous sections may
be related to gauge theories with the help of the abelian projection.
In section 6 we use the methods of Sections 3, 4 to
construct models with exact topological solitons with nontrivial
values of some exotic topological charges like, i.e., $\pi_4(S^3)$ or
$\pi_5(S^4)$.
In Section 7 theories with the more conventional $\pi_n(S^n)$ textures are
investigated. We construct some families of infinitely many finite
energy solutions and further find that they obey a Bogomolny equation.
Finally, we present our conclusions in Section 8.
\section{Generalized integrability}
\subsection{Generalized Zero Curvature (GZC) formulation}
The most natural geometrical object in the generalized zero curvature
representation is a connection on higher loop space, and the condition of
zero curvature for this connection will, in general, not lead to local
equations in ordinary space time. There exist, however, sufficient
local conditions which ensure the vanishing of
the pertinent higher dimensional curvature and, therefore, realize
the generalized zero curvature formulation in a local manner.
One such sufficient condition is constructed as follows.
The starting point is the specification of a Lie algebra
$\mathcal{G}$ and an Abelian ideal $\mathcal{P}$ together with a
connection $A_{\mu}\in \mathcal{G}$ and a rank $d$ antisymmetric tensor
$B_{\mu_1...\mu_d} \in \mathcal{P}$. The
corresponding curvature vanishes if we assume that the connection is
flat and the Hogde dual to $B_{\mu_1...\mu_d}$ is covariantly
constant with respect to the connection i.e.,
\begin{equation}
F_{\mu \nu} (A)=0, \;\;\;\;\; D_{\mu} \tilde{B}^{\mu}=0, \;\;\;\;\;
\mbox{where} \;\;\;\; \tilde{B}_{\mu}\equiv \frac{1}{d!}
\epsilon_{\mu \mu_1...\mu_d} B^{\mu_1...\mu_d}. \label{local def}
\end{equation}
We say that a model possesses the generalized zero curvature
representation if its equations of motion may be re-expressed in this form.
Further, one can notice that for a
given $\tilde{B}_{\mu}$ field we are able to construct conserved
currents which are equal in number to the dimension of the Abelian ideal we
used in the construction. Therefore, we say that a model is
integrable (within this generalized approach) if the corresponding
Abelian ideal has infinite dimensions.
\subsection{Models with 2dim target space}
Let us investigate how this general approach works in the case of
models with two dimensional target space. Here we identify the
target space of the nonlinear model with a two-dimensional manifold
$\mathcal{M}$. Instead of real coordinates $(\xi^1,\xi^2)$ we
introduce the complex coordinates $u=\xi^1+i\xi^2$. According to the
general prescription we fix the Lie algebra and the Abelian ideal.
Namely, $\mathcal{G}$ is the Lie algebra of the $SU(2)$ Lie group
restricted to $S^2$ whereas $\mathcal{P}$ is the representation
space of it with arbitrary angular momentum number and magnetic
number restricted to $\pm 1$, that is $\mathcal{P}=\{\; \mbox{reps}
\; R_{lm} \; \mbox{of} \; su(2), \; m=\pm 1, \; l=1...\infty \}$.
Then, in the triplet representation
\begin{equation}
A_{\mu}=-\partial_{\mu}W W^{-1}=\frac{1}{1+|u|^2} \left( -iu_{\mu}
T_+-i\bar{u}_{\mu}T_- + (u\bar{u}_{\mu} -\bar{u} u_{\mu})T_3 \right)
\end{equation}
\begin{equation}
\tilde{B}_{\mu}=\frac{1}{1+|u|^2} \left(
\bar{\mathcal{H}}_{\mu}P^{(1)}_1- \mathcal{H}_{\mu} P^{(1)}_{-1}
\right),
\end{equation}
where $\mathcal{H}_{\mu}$ is so far an arbitrary vector depending on
the fields as well as their derivatives, $W$ is an element of
$SU(2)/U(1)$ given by
\begin{equation}
W=\frac{1}{\sqrt{1+|u|^2}} \left(
\begin{array}{cc}
1 & iu \\
i\bar{u} & 1
\end{array} \right)
\end{equation}
and $T_3=\mbox{diag} (1,-1)$. Further, $T_{\pm}=(T_1\pm T_2)/2$
and $P^{(j)}_m$ constitute the basis of the Lie algebra and the
Abelian ideal, respectively. $T_1,T_2,T_3$ are Pauli matrices. The
commutators are $[T_3,T_{\pm}]=\pm T_{\pm}, \;\; [T_+,T_-]=2T_3$,
$[T_3,P_m^{(j)}]=m P_m^{(j)}$, $[T_{\pm},P_m^{(j)}] =
\sqrt{j(j+1)-m(m\pm 1)} P_{m \pm 1}^{(j)}$,
$[P^{(j)}_m,P^{(j')}_m]=0$. The connection $A_{\mu}$ is flat by
construction. Thus, the only nontrivial condition in the GZC
formulation is the covariant constancy of the $\tilde{B}_{\mu}$
field. In the triplet representation this results in
\begin{equation}
(1+|u|^2)\partial^{\mu} \mathcal{H}_{\mu}-2u \mathcal{H}_{\mu}
\bar{u}^{\mu}=0. \label{2dim int eq gen}
\end{equation}
However, in a higher spin representation one gets, in addition to
(\ref{2dim int eq gen}), the constraint
\begin{equation}
\mathcal{H}_{\mu} \bar{u}^{\mu}=0. \label{2dim constrain}
\end{equation}
So, we can conclude that a dynamical model with two dimensional
target space is integrable if one may define a vector quantity
$\mathcal{H}_{\mu}$ such that $\mathcal{H}_{\mu}\bar{u}^{\mu}\equiv
0 $ and the pertinent equations of motion read
\begin{equation}
\partial^{\mu} \mathcal{H}_{\mu}=0. \label{2dim int eq}
\end{equation}
Models with these properties are known as models of the
Aratyn-Ferriera-Zimmerman type \cite{afz2}. They are integrable in
the GZC formulation sense: they have the GZC formulation with the
infinite-dimensional Abelian ideal. They are given by the
following Lagrange density
\begin{equation}
\mathcal{L}= \omega(u\bar{u})  H^q,
\end{equation}
where
\begin{equation}
H \equiv  u_{\mu}^2 \bar{u}_{\nu}^2 - (u_{\mu} \bar{u}^{\mu})^2.
\label{H afz}
\end{equation}
$\omega$ is any function of $u\bar{u}$ whereas $q$ is a positive
real parameter. A particular example of such integrable models in
four dimensional Minkowski space-time is given by the expression
\begin{equation}
\mathcal{L}_{AFZ}= \omega(u\bar{u}) H^{\frac{3}{4}}, \label{afz}
\end{equation}
where the value of the power is taken to avoid the Derrick
arguments for the non-existence of static solitons \cite{afz1}.
The AFZ model describes, in fact, soliton excitations of a three
component unit vector field which may be related via the standard
stereographic projection with the complex field $u$. As the static
solutions are maps from compactified $R^3$ to the $S^2$ target
space they carry the corresponding topological charge, i.e., the Hopf
index $Q
\in \pi_3(S^2) \cong Z$. The lump like structure of the solitons
emerges from the fact that the pre-image of a given point on the
target sphere is a closed line. For this model such topologically
nontrivial solitons (hopfions) have been derived in an exact form
\cite{afz2}. Moreover, one can also construct infinitely many
conserved currents
\begin{equation}
j_{\mu}= G_{\bar{u} } \mathcal{H}_{\mu}- G_{u }
\bar{\mathcal{H}}_{\mu}, \label{afz currents}
\end{equation}
where
\begin{equation}
\mathcal{H}_{\mu}=\omega^{1/3} H^{-1/4} h_{\mu} \;\;\; \mbox{and}
\;\;\; h_{\mu}\equiv H_{u^{\mu}}=2  \left( \bar{u}_{\nu}^2 u_{\mu} -
(u_{\nu} \bar{u}^{\nu}) \bar{u}_{\mu}\right).
\end{equation}
Further, $G$ is an arbitrary function of $u$ and $\bar u$, and
$G_u \equiv \partial_u G$, etc.
In order to understand the geometrical meaning of this model, which
may give us a clue how to generalize it to field theories with a
more complicated target space, we consider the area two form on the
target space manifold
\begin{equation}
\Omega \equiv \frac{g(u\bar{u})}{2i} d\bar{u} \wedge du, \label{area
n2}
\end{equation}
where $g$ is the area density. The pullback of the area two-form in the
base $(d+1)$ Minkowski space-time is
\begin{equation}
\Omega' = \frac{g(u\bar{u})}{2i} \bar{u}_{\mu} u_{\nu} dx^{\mu}
\wedge dx^{\nu}. \label{pull area n2}
\end{equation}
Now we are able to define two objects. Namely, a rank two
anti-symmetric tensor
\begin{equation}
h_{\mu \nu} \equiv u_{\mu}\bar{u}_{\nu}-\bar{u}_{\mu}u_{\nu}
\label{h afz}
\end{equation}
and a scalar density $H \equiv \frac{1}{2!} h_{\mu \nu}^2$ which are
exactly the same objects as used in the construction of the AFZ
model. Observe that the density $H$ equals the square of the
pullback of the area two-form modulo a multiplicative term depending
on the area density $g$. In other words, the fact which makes the
AFZ model integrable is that it is proportional to a function of
the square of the pullback of the area two-form to the base Minkowski
space-time.
\\
Therefore, one can conjecture that integrable models with higher
dimensional target spaces can be constructed using the square of the
pullback of the pertinent volume form on the target space manifold
into the base Minkowski space-time. In the proceeding sections
we demonstrate this hypothesis by explicit construction.
\section{Integrable models with 3dim target space}
\subsection{The model}
Following the considerations of the previous section, our starting point
for the construction of integrable models with three
dimensional target space $\mathcal{M}^{(3)}$ is to consider the
volume three-form on $\mathcal{M}^{(3)}$
\begin{equation}
V_{(3)}= \frac{g(u\bar{u}, \xi)}{2i} du \wedge d\bar{u} \wedge
d\xi, \label{vol n3}
\end{equation}
where $u,\bar{u}$ together with a scalar $\xi$ are local
coordinates on $\mathcal{M}^{(3)}$ and $g$ is the volume density.
Then the pullback of the volume three-form to the base Minkowski
space-time is
\begin{equation}
V'_{(3)}= \frac{g(u\bar{u}, \xi)}{2i}
u_{\mu}\bar{u}_{\nu}\xi_{\rho} dx^{\mu} \wedge dx^{\nu} \wedge
dx^{\rho}. \label{pull vol n3}
 \end{equation}
Again, we may extract from the last formula a rank three
anti-symmetric tensor
\begin{equation}
h_{\mu \nu \rho}= u_{\mu}\bar{u}_{\nu}\xi_{\rho} +
u_{\rho}\bar{u}_{\mu}\xi_{\nu} +u_{\nu}\bar{u}_{\rho}\xi_{\mu} -
u_{\nu}\bar{u}_{\mu}\xi_{\rho}-u_{\rho}\bar{u}_{\nu}\xi_{\mu}-
u_{\mu}\bar{u}_{\rho}\xi_{\nu}
\label{h n3}
\end{equation}
and the corresponding scalar
\begin{equation}
H_{(3)} \equiv \frac{1}{3!} h_{\mu \nu \rho}^2 = \xi^2_{\rho}
\left(u_{\mu}^2 \bar{u}_{\nu}^2 - (u_{\mu} \bar{u}^{\mu})^2
\right) + 2 (u_{\mu} \bar{u}^{\mu}) (u_{\nu} \xi^{\nu})
(\bar{u}_{\rho} \xi^{\rho}) - (u_{\mu}\xi^{\mu})^2\bar{u}_{\nu}^2-
(\bar{u}_{\mu}\xi^{\mu})^2 u_{\nu}^2. \label{H n3}
\end{equation}
The last object is proportional to the square of the pullback of
the volume three-form up to a term which does not contain any
derivatives of the fields. \footnote{Such a rank three tensor in a
slightly different parametrization has been previously analyzed in
the context of the so-called generalization of the Goldstone model
in (3+1) dimensions, which solutions are ungauged Higgs analogues
of the Skyrme model solitons \cite{radu1}.}
\\
Then, the class of integrable models with tree dimensional target
space is defined as follows
\begin{equation}
\mathcal{L}=\omega(u\bar{u}, \xi) H_{(3)}^q, \label{model n3}
\end{equation}
with a positive parameter $q$ (where $q$ may, e.g., be chosen to guarantee the
invariance of the model under the scale transformation or to
provide finite energy solutions.)
\\
In order to write the corresponding equations of motion let us
define two vector quantities closely related to the canonical
momenta
$$
h_{\mu} \equiv \frac{\partial H_{(3)}}{\partial u^{\mu}} = $$
\begin{equation}
2 \xi^2 \left( \bar{u}_{\nu}^2 u_{\mu} - (u_{\nu} \bar{u}^{\nu})
\bar{u}_{\mu}\right) +2(u_{\nu} \xi^{\nu}) (\bar{u}_{\nu} \xi^{\nu})
\bar{u}_{\mu} + 2(u_{\nu} \bar{u}^{\nu}) (\bar{u}_{\nu} \xi^{\nu})
\xi_{\mu} - 2 (u_{\nu} \xi^{\nu})\bar{u}_{\rho}^2 \xi_{\mu} -  2
(\bar{u}_{\mu}\xi^{\mu})^2 u_{\mu} \label{vec h n3}
\end{equation}
and
\begin{equation}
k_{\mu} \equiv \frac{\partial H_{(3)}}{\partial \xi^{\mu}} = 2
\left(u_{\mu}^2 \bar{u}_{\nu}^2 - (u_{\nu} \bar{u}^{\nu})^2
\right) \xi_{\mu} + 2 (u_{\nu} \bar{u}^{\nu}) \left( (u_{\nu}
\xi^{\nu}) \bar{u}_{\mu} + (\bar{u}_{\nu} \xi^{\nu}) u_{\mu}
\right) - 2 (u_{\nu} \xi^{\nu}) \bar{u}_{\nu}^2 u_{\mu} -
2(\bar{u}_{\nu} \xi^{\nu}) u_{\nu}^2 \bar{u}_{\mu}. \label{vec k
n3}
\end{equation}
It is easy to verify that they obey the following relations
\begin{equation}
h_{\mu} u^{\mu} = 2H_{(3)} \label{vec h n3 prop1}
\end{equation}
\begin{equation}
h_{\mu} \bar{u}^{\mu} =0, \;\;\;\;\; h_{\mu} \xi^{\mu}=0
\label{vec h n3 prop2}
\end{equation}
and
\begin{equation}
k_{\mu} \xi^{\mu}=2H_{(3)} \label{vec k n3 prop1}
\end{equation}
\begin{equation}
k_{\mu} u^{\mu}=0, \;\;\;\;\;\; k_{\mu} \bar{u}^{\mu}=0 \label{vec
k n3 prop2}
\end{equation}
Therefore, the equations of motion
\begin{equation}
\partial_{\mu} \left( q \omega H_{(3)}^{q-1}
h^{\mu}\right)- \omega'\bar{u}H_{(3)}^{q}=0 \label{eom h n3}
\end{equation}
\begin{equation}
\partial_{\mu} \left( q \omega H_{(3)}^{q-1}
k^{\mu}\right)- \omega_{\xi} H_{(3)}^{q}=0 \label{eom k n3}
\end{equation}
may be rewritten in the following simple form
\begin{equation}
\partial_{\mu} \mathcal{H}^{\mu}=0, \;\;\;\;\; \mathcal{H}_{\mu}
\equiv \omega^{1-\frac{1}{2q}} H_{(3)}^{q-1} h_{\mu}, \label{eom h
n3 1}
\end{equation}
\begin{equation}
\partial_{\mu} \mathcal{K}^{\mu}=0, \;\;\;\;\; \mathcal{K}_{\mu}
\equiv \omega^{1-\frac{1}{2q}} H_{(3)}^{q-1} k_{\mu}. \label{eom k
n3 1}
\end{equation}
The prime denotes differentiation with respect to $u\bar{u}$.
\subsection{GZC formulation}
To express the system of equations (\ref{eom h n3 1}), (\ref{eom k
n3 1}) in terms of the generalized zero curvature condition, we
specify $\mathcal{G}$ to be the Lie algebra of the $SU(2)$ Lie
group (restricted to the equator) while $\mathcal{P}$ is the
representation space of it with arbitrary integer angular momentum
quantum number $l$, but magnetic quantum number $m$ restricted to
$\pm 1, 0$. As one can see, there is only one change in comparison
with the integrable models with two dimensional target space. The
Abelian ideal is extended to those representations which carry
also zero magnetic number. Then, in spin $(j)$ representation, the
flat connection and the Hodge dual field are
\begin{equation}
A_{\mu} = -\partial_{\mu} W \; W^{\dagger} = \frac{1}{1+|u|^2}
\left( -iu_{\mu} T_+ - i\bar{u}_{\mu} T_- + (u
\bar{u}_{\mu}-\bar{u}u_{\mu} ) T_3 \right)
\end{equation}
and
\begin{equation}
\tilde{B}_{\mu}^{(j)} = \frac{i}{(1+|u|^2)^2} \mathcal{K}_{\mu}
P^{(j)}_0+\frac{1}{1+|u|^2} \left(\bar{\mathcal{H}}_{\mu}
P^{(j)}_1 - \mathcal{H}_{\mu} P^{(j)}_{-1} \right).
\end{equation}
Similar fields are also used in the GZC formulation of the Skyrme
model, see \cite{joaquin S3}. The covariant constancy of the Hodge
dual field $\tilde{B}_{\mu}$ gives
$$
\frac{i}{(1+|u|^2)^2} \partial^{\mu} \mathcal{K}_{\mu} P^{(j)}_0 +
\frac{1}{1+|u|^2} \left(\partial^{\mu}\bar{\mathcal{H}}_{\mu}
P^{(j)}_1 - \partial^{\mu} \mathcal{H}_{\mu} P^{(j)}_{-1} \right)
- \frac{1}{(1+|u|^2)^2} \left(u\bar{u}^{\mu}
\bar{\mathcal{H}}_{\mu} P^{(j)}_1 - \bar{u} u^{\mu}
\mathcal{H}_{\mu} P^{(j)}_{-1} \right)-
$$
\begin{equation}
\frac{i}{(1+|u|^2)^2} \left( \bar{u}^{\mu}
\bar{\mathcal{H}}_{\mu} -u^{\mu} \mathcal{H}_{\mu}\right)
\sqrt{j(j+1)} P^{(j)}_0 +\frac{1}{(1+|u|^2)^2} u\bar{u}^{\mu}
\bar{\mathcal{H}}_{\mu} P^{(j)}_1  - \frac{1}{(1+|u|^2)^2}\bar{u}
u^{\mu} \mathcal{H}_{\mu} P^{(j)}_{-1}=0.
\end{equation}
Here we used the following important properties obeyed by the
objects $\mathcal{K}_{\mu}$ and $\mathcal{H}_{\mu}$. Namely,
\begin{equation}
\mathcal{H}_{\mu} \bar{u}^{\mu}=0, \;\; \mathcal{H}_{\mu}
\xi^{\mu}=0, \;\;\;\; \mathcal{K}_{\mu} u^{\mu}=0, \;\;
\mathcal{K}_{\mu} \bar{u}^{\mu}=0.
\end{equation}
Moreover, if we notice that
\begin{equation}
\mathcal{H}_{\mu} u^{\mu}= \bar{u}^{\mu} \bar{\mathcal{H}}_{\mu}
\end{equation}
then we arrive at the field equations (\ref{eom h n3 1}),
(\ref{eom k n3 1}). Therefore, we conclude that these models are
integrable. The Abelian ideal we used in the generalized zero
curvature is indeed infinite dimensional.
\\
Observe that the connection $A_\mu$ belongs to the Lie algebra of
the $SU(2)$ Lie group restricted to the coset space $SU(2)/U(1)$,
as is the case for models with two-dimensional target space.
Moreover, the dual field $\tilde{B}_{\mu}$ is defined up to an
arbitrary function of $u$ and $\bar{u}$ which multiplies
$P^{(j)}_0$.
\\
Finally, we prove that this family of models possess infinitely many
conserved quantities as is required for the integrable systems.
After some calculations one can verify that there are three families
of infinitely many on-shell conserved currents
\begin{equation}
j_{\mu}^{(G)} = G_{\bar{u}} \mathcal{H}_{\mu} -G_{u}
\bar{\mathcal{H}}_{\mu} \label{curr G n3 1}
\end{equation}
\begin{equation}
j_{\mu}^{(\tilde G)} = \tilde G_{\xi} \mathcal{H}_{\mu}- \tilde G_{u}
\mathcal{K}_{\mu} \label{curr G n3 2}
\end{equation}
\begin{equation}
j_{\mu}^{(\tilde {\tilde G})} = \tilde {\tilde G}_{\xi}
\bar{\mathcal{H}}_{\mu}
- \tilde {\tilde G}_{\bar{u}}
\mathcal{K}_{\mu}. \label{curr G n3 3}
\end{equation}
Here
\begin{equation}
G=G(u,\bar{u},\xi), \;\;\; \tilde G=\tilde G(u,\bar{u},\xi),
\;\;\; \tilde {\tilde G} = \tilde {\tilde G} (u,\bar{u},\xi),
\;\;\;. \label{curr dep n3}
\end{equation}
Moreover, there is a good understanding of the geometrical origin of
the currents. We show that the conservation laws found for the
integrable models are generated by a class of geometric target space
transformations. Specifically, they are the Noether currents related
to the volume-preserving diffeomorphisms. Let us again consider a
three-dimensional target space manifold $\mathcal{M}^{(3)}$,
parameterized by local coordinates $X^i, i=1,2,3$. Then the volume
3-form is given by the expression
\begin{equation}
V_{(3)}= g(X^i) dX^1 \wedge dX^2 \wedge dX^3. \label{vol n3 a}
\end{equation}
A volume-preserving diffeomorphism is a a coordinate
transformation leaving the volume form invariant. For an arbitrary
infinitesimal transformation
\begin{equation}
X^{'i}=X^i+\epsilon Y^i(X^j)
\end{equation}
invariance of the volume form results in the condition on $Y^i$
functions
\begin{equation}
\partial_i(gY^i)\equiv \frac{\partial}{\partial X^i} (g Y^i)=0.
\end{equation}
As the considered manifold is three dimensional we apply Darboux's
theorem and derive a general (local) solution
\begin{equation}
gY^i=\epsilon^{ijk}\partial_j A \partial_k B,
\end{equation}
where $A,B$ are arbitrary functions of the local coordinates.
Following that we may write a general vector field generated by a
volume-preserving diffeomorphisms
\begin{equation}
\mathbf{v}^{Y}=Y^i \partial_i= Y^u\partial_u+Y^{\bar{u}}
\partial_{\bar{u}}+Y^{\xi}\partial_{\xi},
\end{equation}
where we assumed the parametrization of the target
manifold by a complex field $u$ and a real scalar $\xi$ introduced before.
These vector fields obey the Lie algebra
\begin{equation}
[\mathbf{v}^{Y},\mathbf{v}^{\tilde{Y}}]=\mathbf{v}^{\tilde{\tilde{Y}}},
\end{equation}
\begin{equation}
\tilde{\tilde{Y}}^i=(\partial_j Y^i) \tilde{Y}^j-(\partial_j
\tilde{Y}^i) Y^j.
\end{equation}
In a relativistic field theory one can find a general expression
for Noether currents corresponding to the vector fields
$\mathbf{v}^{Y}$. Namely,
\begin{equation}
J^{(Y)}_{\mu}=Y^u\pi_{\mu}+Y^{\bar{u}} \bar{\pi}_{\mu}+Y^{\xi}
P_{\mu},
\end{equation}
where $\pi_{\mu}=\partial_{u_{\mu}} \mathcal{L},
\bar{\pi}_{\mu}=\partial_{\bar{u}_{\mu}} \mathcal{L},
P_{\mu}=\partial_{\xi_{\mu}} \mathcal{L}$ are the standard
canonical momenta. For the integrable models discussed in the
previous section we get (up to an unimportant multiplicative
constant)
\begin{equation}
\pi_{\mu} = \omega^{\frac{1}{2q}} \mathcal{H}_{\mu}, \;\;\;
\bar{\pi}_{\mu} = \omega^{\frac{1}{2q}} \bar{\mathcal{H}}_{\mu},
\;\;\; P_{\mu} = \omega^{\frac{1}{2q}} \mathcal{K}_{\mu}.
\end{equation}
\section{Higher dimensional target space}
The generalization to integrable field theories with a target space of
arbitrary dimension $n$ is straightforward. As we described in the
case of models with two or three dimensional target spaces, one
should begin with the pertinent volume form on the target space
manifold $\mathcal{M}^{(n)}$ i.e.,
\begin{equation}
V_{(2k)}=g(u^{(i)},\bar{u}^{(i)}) du^{(1)}\wedge
d\bar{u}^{(1)}\wedge ...\wedge du^{(k)}\wedge d\bar{u}^{(k)}
\end{equation}
for an even number of dimensions $n=2k$ or
\begin{equation}
V_{(2k+1)}=g(u^{(i)},\bar{u}^{(i)}, \xi) du^{(1)}\wedge
d\bar{u}^{(1)}\wedge ...\wedge du^{(k)}\wedge d\bar{u}^{(k)}\wedge
d\xi
\end{equation}
for an odd number of dimensions $n=2k+1$. The local coordinates on
$\mathcal{M}$ are $u^{(1)}, \bar{u}^{(1)},...,u^{(k)},
\bar{u}^{(k)}$ or $u^{(1)}, \bar{u}^{(1)},...,u^{(k)},
\bar{u}^{(k)}$, $\xi$ respectively for those cases. $u^{(i)},
i=1...k$ are complex fields while $\xi$ is a real scalar. Therefore,
after performing the pullback into the base Minkowski space-time we
can define a rank $n$ antisymmetric tensor $h_{\mu_1...\mu_n}$ as
\begin{equation}
h_{\mu_1...\mu_{2k}}=u^{(1)}_{[\mu_1}
\bar{u}^{(1)}_{\mu_2}...u^{(k)}_{\mu_{2k-1}}
\bar{u}^{(k)}_{\mu_{2k}]} \;\;\; \mbox{or} \;\;\;
h_{\mu_1...\mu_{2k+1}}=u^{(1)}_{[\mu_1}
\bar{u}^{(1)}_{\mu_2}...u^{(k)}_{\mu_{2k-1}}
\bar{u}^{(k)}_{\mu_{2k}} \xi_{\mu_{2k+1}]}
\end{equation}
depending on the dimension of $\mathcal{M}$. Here $[\;]$ stands
for antisymmetrization. In addition we need a scalar quantity
built out of this tensor
\begin{equation}
H_{(n)}=\frac{1}{n!} h_{\mu_1...\mu_n}^2.
\end{equation}
The integrable Lagrangian reads
\begin{equation}
\mathcal{L}_{(n)}= \omega H^q_{(n)},
\end{equation}
where $\omega $ is a real function factor depending on the fields
i.e., coordinates on the n-dimensional manifold
$\mathcal{M}^{(n)}$. Then the equations of motion read
\begin{equation}
\partial^{\mu} \mathcal{H}^{(i)}_{\mu}=0, \;\;\;\;\;
\mathcal{H}_{\mu}^{(i)} \equiv \omega^{1-\frac{1}{2q}} H_{(n)}^{q-1}
h_{\mu}^{(i)}, \;\;\;i=1...k \label{eom h n}
\end{equation}
and their complex conjugate together with
\begin{equation}
\partial_{\mu} \mathcal{K}^{\mu}=0, \;\;\;\;\; \mathcal{K}_{\mu}
\equiv \omega^{1-\frac{1}{2q}} H_{(n)}^{q-1} k_{\mu} \label{eom k n}
\end{equation}
if the dimension of the target space is odd. Here
\begin{equation}
h_{\mu}^{(i)} \equiv \frac{\partial H_{(n)}}{\partial u^{(i) \,
\mu}}, \;\;\;\; i=1..k \;\;\;\;\; \mbox{and} \;\;\;\;\; k_{\mu}
\equiv \frac{\partial H_{(n)}}{\partial \xi^{\mu}}
\end{equation}
are the canonical momenta obeying the following constrains
identically
\begin{equation}
h_{\mu}^{(i)} u^{(i)\mu}=2H_{(n)}, \;\;\;\;\; \mbox{and} \;\;\;\;\;
k_{\mu} \xi^{\mu}=2H_{(n)} \;\;\;\;\; \mbox{no sumation over} \;\;
i=1...k
\end{equation}
\begin{equation}
h_{\mu}^{(i)} u^{(j)\mu}=h_{\mu}^{(i)} \xi^{\mu}=0, \;\;\;\;
\mbox{for all} \;\;\;\; i \neq j
\end{equation}
\begin{equation}
h_{\mu}^{(i)} \bar{u}^{(j)\mu}=0 \;\;\;\;\; \mbox{for all} \;\;\;\;
i,j.
\end{equation}
\begin{equation}
k_{\mu} u^{(j)\mu}=k_{\mu} \bar{u}^{(j)\mu}=0 \;\;\;\;\; \mbox{for
all} \;\;\;\; i,j.
\end{equation}
The families of infinitely many conserved quantities are given by
the formulas
\begin{equation}
j_{\mu}^{(i,j)}=G^{(i,j)}_{\bar{u}_j}
\mathcal{H}^{(i)}_{\mu}-G^{(i,j)}_{u_i}
\bar{\mathcal{H}}^{(j)}_{\mu}, \;\;\;\; i,j=1...k  \label{cur gen
1}
\end{equation}
\begin{equation}
\tilde{j}_{\mu}^{(i,j)}=\tilde{G}^{(i,j)}_{\bar{u}_j}
\mathcal{H}^{(i)}_{\mu}-\tilde{G}^{(i,j)}_{\bar{u}_i}
\mathcal{H}^{(j)}_{\mu}, \;\;\;\;
\tilde{\tilde{j}}_{\mu}^{(i,j)}=\tilde{\tilde{G}}^{(i,j)}_{u_j}
\bar{\mathcal{H}}^{(i)}_{\mu}-\tilde{\tilde{G}}^{(i,j)}_{u_i}
\bar{\mathcal{H}}^{(j)}_{\mu}, \;\;\;\; i,j=1...k \label{cur gen
2}
\end{equation}
\begin{equation}
j^{(i)}_{\mu}=G^{(i)}_{\xi} \mathcal{H}^{(i)}_{\mu} -
G^{(i)}_{u_i}\mathcal{K}_{\mu}, \;\;\;\;
\tilde{j}^{(i)}_{\mu}=\tilde{G}^{(i)}_{\xi}
\bar{\mathcal{H}}^{(i)}_{\mu} -
\tilde{G}^{(i)}_{\bar{u}_i}\mathcal{K}_{\mu}, \;\;\;\; i=1...k,
\label{cur gen 3}
\end{equation}
where all functions $G$ are arbitrary functions of all scalar fields
and there is no summation over the $i,j$ indices. Thus, in the case
of even dimensional target space $n=2k$ we have
$k^2+2(k^2-k)=k(3k-2)$ independent families of infinitely many
conserved currents. For odd dimensions $n=2k+1$ this number is
$3k^2$.

\vspace*{0.2cm}

\noindent {\bf Remark:}
\\
When the dimension of the target space $n$ equals the
number of the spatial dimension of the base space $d$, then the
static equations of motion are trivial. In fact, then the quantity
$h_{\mu_1...\mu_n} dx^{\mu_1}\wedge...\wedge dx^{\mu_n}$, being the pullback of
the volume form on a target space, is a closed $n$-form also in
the base space. As a consequence, its integral over a
$n$-dimensional manifold only depends on the boundary conditions,
that is, it gives a purely topological action (or energy
functional). Therefore, these integrable models do not seem to lead to
interesting results in the context of $\pi_n(S^n)$ solitons. (In
Section 7 we show how one can circumvent this obstacle and
construct actions with solitons of $\pi_n(S^n)$ type. The main
idea is to include several pullback tensors in the Lagrangian.) If
$d>n$, then the models are not longer trivial and may provide
exact soliton solutions with topological charges from the
homotopy group $\pi_d(S^n)$,
as has been observed in the original AFZ model.
\section{Relation to gauge theories}
It has been established that two dimensional target space
integrable models of the AFZ type may emerge via the Abelian projection
of the $SU(2)$ Yang-Mills field. It results in the
observation that the Abelian projection of $SU(2)$ YM is
an integrable sector of the full theory with magnetic monopoles as
exact solutions \cite{we YMdil}. In the case of integrable
models with higher dimensional target space the situation is
analogous. There exists a gauge theory which, after reduction of
degrees of freedom, leads to the corresponding integrable pullback model.
\\
As a first example we consider the following model
\begin{equation}
L=\xi_{\rho}^2 F^a_{\mu \nu} F^{a \mu \nu} - 2 (F^a_{\mu \nu}
\xi^{\nu})^2, \label{gauge 3d}
\end{equation}
where the gauge fields $A_{\mu}^a, a=1,2,3$ from the $su(2)$ Lie
algebra is non-minimally coupled to the scalar field $\xi$. The
field strength tensor is defined in the standard manner $F^a_{\mu
\nu}=\partial_{\mu}A^a_{\nu} - \partial_{\nu} A^a_{\mu} +
\epsilon^{abc}A^b_{\mu}A^c_{\nu}$. The next step is to use the
Cho-Faddeev-Niemi-Shabanov decomposition \cite{cho1}-\cite{gies}
and express the gauge fields by means of a new set of degrees of
freedom
\begin{equation}
A^a_{\mu}=C_{\mu}n^a+\epsilon^{abc}n^b_{\mu}n^c + W^a_{\mu},
\label{decomp}
\end{equation}
where we introduced a three component unit vector field $\vec{n}$
pointing into the color direction, an Abelian gauge potential
$C_{\mu}$ and a color vector field $W_{\mu}^a$ which is
perpendicular to $\vec{n}$. Now, we restrict the gauge potential
to the form
\begin{equation}
A^a_{\mu}=\epsilon^{abc}n^b_{\mu}n^c.
\end{equation}
Then the field strength tensor reads
\begin{equation}
F^a_{\mu \nu}=-\epsilon^{abc}n^b_{\mu} n^c_{\nu}.
\end{equation}
Finally, taking into account the stereographic projection
\begin{equation}
\vec{n}= \frac{1}{1+|u|^2} \left( u+u^*, -i(u-u^*), |u|^2-1
\right) \label{stereo}
\end{equation}
we arrive at the Lagrange density
\begin{equation}
L=-\frac{8}{(1+|u|^2)^4} \left[ \xi^2_{\rho} \left(u_{\mu}^2
\bar{u}_{\nu}^2 - (u_{\mu} \bar{u}^{\mu})^2 \right) + 2 (u_{\mu}
\bar{u}^{\mu}) (u_{\nu} \xi^{\nu}) (\bar{u}_{\rho} \xi^{\rho}) -
(u_{\mu}\xi^{\mu})^2\bar{u}_{\nu}^2- (\bar{u}_{\mu}\xi^{\mu})^2
u_{\nu}^2 \right].
\end{equation}
As we claimed, it has exactly the form of the integrable pullback
model with three-dimensional target space (\ref{model n3}), with
$q=1$ and $\omega =8/(1+|u|^2)^4$.
\\
In the case of the four-dimensional integrable models the related
$SU(2)\times SU(2)$ gauge theory reads
\begin{equation}
L=(F^a_{\mu \nu})^2 (G^b_{\rho \sigma})^2-4 F^a_{\mu \rho} F^{a
\mu \nu} G^b_{\sigma \nu} G^{b \sigma \rho} + F^a_{\mu \nu}
F^a_{\rho \sigma} G^{b \mu \nu} G^{b \rho \sigma}, \label{gauge
4d}
\end{equation}
where we have introduced the second, independent $SU(2)$ gauge
potential $B^a_{\mu}$. Then, $G^a_{\mu
\nu}=\partial_{\mu}B^a_{\nu} - \partial_{\nu} B^a_{\mu} +
\epsilon^{abc}B^b_{\mu}B^c_{\nu}$ and the corresponding
abelian projection is performed assuming
\begin{equation}
A^a_{\mu}=\epsilon^{abc}n^b_{\mu}n^c, \;\;\;\;
B^a_{\mu}=\epsilon^{abc}m^b_{\mu}m^c,
\end{equation}
where $\vec{m}$ is another unit, three component vector field. Of
course, one may continue this procedure and try to find gauge
theories related to higher dimensional integrable pullback models.
However, we would like to have a constructive method for
generating such models, instead of this guessing-like procedure.
Fortunately, such a method exists and is based on the observation
that the gauge models (\ref{gauge 3d}) and (\ref{gauge 4d}) can be
written in a more compact form. Namely, the Lagrange density
(\ref{gauge 3d}) is given by
\begin{equation}
L= j^a_{\mu_1...\mu_{d-2}}j^{a \mu_1...\mu_{d-2}},
\end{equation}
where
\begin{equation}
j^a_{\mu_1...\mu_{d-2}}=\epsilon_{\mu_1...\mu_{d-2} \mu \nu \rho}
F^{a \mu \nu} \xi^{\rho}.
\end{equation}
In the same way, the model (\ref{gauge 4d}) may be expressed as
\begin{equation}
L= j^{a b}_{\mu_1...\mu_{d-3}}j^{a b \mu_1...\mu_{d-3}},
\end{equation}
where
\begin{equation}
j^{ab}_{\mu_1...\mu_{d-3}}=\epsilon_{\mu_1...\mu_{d-3} \mu \nu
\rho \sigma} F^{a \mu \nu} G^{b \rho \sigma}.
\end{equation}
$d$ is the number of the spatial dimensions. Now, we are able to
define gauge models which can be reduced via the Abelian
projection to the integrable pullback models. Equivalently, one
can say that these non-integrable gauge theories possess an
integrable sector given by the pertinent integrable pullback
model. The specific Lagrange density reads
\begin{equation}
L=j^{a_1..a_p}_{\mu_{2p+1}...\mu_{d+1}}j^{a_1..a_p \; \mu_{2p+1}...\mu_{d+1}}
\end{equation}
where $2p$ is an even dimension of the target space while $d$ is
the number of the spatial dimensions. Here
\begin{equation}
j^{a_1..a_p}_{\mu_{2p+1}...\mu{d+1}}=\epsilon_{\mu_1...\mu_{2p-1}
\mu_{2p}...\mu_{d+1}}F_{(1)}^{a_1 \mu_1 \mu_2}...\; F_{(p)}^{a_p
\mu_{2p-1} \mu_{2p}}.
\end{equation}
If the target space has an odd dimension $2p+1$, then we find
\begin{equation}
L=j^{a_1..a_p}_{\mu_{2p+2}...\mu_{d+1}}j^{a_1..a_p \; \mu_{2p+2}...\mu_{d+1}}
\end{equation}
and
\begin{equation}
j^{a_1..a_p}_{\mu_{2p+2}...\mu{d+1}}=\epsilon_{\mu_1...\mu_{2p-1}
\mu_{2p} \mu_{2p+1}...\mu_{d+1}}F_{(1)}^{a_1 \mu_1 \mu_2}...\;
F_{(p)}^{a_p \mu_{2p-1} \mu_{2p}}\xi^{\mu_{2p+1}}.
\end{equation}
Here each $F_{(i)}^{a \mu \nu}$ is the $SU(2)$ field strength
tensor defined by an independent $SU(2)$ gauge field $A_{(i)}^{a
\mu}$.
\\
To conclude, we have found $SU(2)\times SU(2) \times ... \times
SU(2)$ theories which give a gauge covering of the integrable
pullback models. The question whether the pullback models may be
immersed in gauge theories of a different type is still an open
problem, which definitely requires further studies.

\vspace*{0.2cm}

\noindent {\bf Remark:}
\\
One might conjecture that the rank three tensor
$h_{\mu \nu \sigma}$ of Section 3 is related by means of some projection
(reduction of degrees of freedom) to a rank three field strength
tensor $\hat{F}_{\mu \nu \rho}$ of a non-Abelian rank two
gauge field $\hat{A}_{\mu \nu}$ in the same way that the rank two tensor
$h_{\mu\nu}$ is related to the abelian projection of an SU(2) field
strength tensor.
This would then imply
that there existed an integrable subsector of the corresponding
higher rank nonabelian gauge theory defined by such a projection.
Unfortunately, the issue of nonabelian higher rank gauge theories
is a difficult one, and a satisfactory definition of these theories
has not yet been found. On the other hand, abelian higher rank gauge theories
cf. Kalb--Ramond theories  \cite{kalb} do exist and have already
demonstrated their relevance in numerous applications. Here we just want to
mention that in $(d+1)$ Minkowski space-time,
our tensor obeys the analog of
the Bianchi identity
$$
\epsilon_{\mu_1 \mu_2 \mu_3 \mu_4...\mu_{d+1}}
\partial^{\mu_1} h^{\mu_2 \mu_3 \mu_4}=0
$$
as is required for the
Abelian Kalb--Ramond tensor
 \cite{kalb mon}
$$
F_{\mu \nu \sigma}=\partial_{\mu}A_{\nu
\sigma}+\partial_{\sigma}A_{\mu \nu}+\partial_{\nu}A_{\sigma
\mu} .
$$
So there may exist embeddings of our pullback model into
 Kalb--Ramond theories, although perhaps not in such a natural way like
the abelian projection in the case of a nonabelian gauge theory.
\section{Examples of exotic textures}
Experience from integrable models in $(1+1)$
dimensions tells us that, in addition to the existence of infinitely
many conservation laws and the zero curvature representation, such
theories allow for exact (soliton) solutions. In this section we
prove that the integrable models described before possess exact
topologically nontrivial solutions. In fact, as our construction is
valid for target and base spaces of any dimension, we are able
to find Lorentz invariant dynamical systems which provide exact
textures carrying rather exotic topological charges.
Higher-dimensional textures with unusual topological charges have already been
investigate, although not from the integrability point of view. Textures for
higher Hopf maps have, e.g., been studied in  \cite{vachaspati}.
A typical field of applications of these higher-dimensional textures is
cosmology, therefore there exist many studies of textures with gravitational
backreaction (i.e., self-gravitating textures). Some examples of the latter are
gravitating 5-dim solitons
\cite{rahaman}, $O(4)$ gravitating solitons \cite{i cho}, gravitating
monopoles in higher extra dim \cite{vilenkin1}, \cite{vilenkin2}, or a
gravitating 6-dim Abelian Higgs vortex \cite{torrealba}.
\\

\subsection{Suspended Hopf maps on $S^4$}
As we mentioned before, the AFZ model possesses soliton
solutions with nontrivial values of the Hopf index $Q_H \in
\pi_3(S^2)$. However, one can associated with these Hopf maps
$\vec{n}: S^3 \rightarrow S^2$ a suspended map $ \vec{N}: S^4
\rightarrow S^3$ by mapping the $S^3$ equator of $S^4$ onto the
$S^2$ equator of $S^3$ using the Hopf map $\vec{n}$ and then
continuing smoothly to the poles. It is known that those new
suspended maps may be classified by a topological invariant as the
pertinent homotopy group is nonzero, $\pi_4(S^3) \cong \ZZ_2$.
Examples of the nontrivial representative class are the suspended
maps
\begin{equation}
\vec{N}: \left(
\begin{array}{c}
R_0\sqrt{z} \cos \phi_2 \sin \eta \\
R_0\sqrt{z} \sin \phi_2 \sin \eta \\
R_0\sqrt{1-z} \cos \phi_1 \sin \eta \\
R_0\sqrt{1-z} \sin \phi_1 \sin \eta \\
R_0\cos \eta
\end{array} \right) \longrightarrow \left(
\begin{array}{c}
n_1 \sin \eta \\
n_2 \sin \eta \\
n_3 \sin \eta \\
\cos \eta
\end{array} \right),
\end{equation}
where $z \in [0,1], \phi_1 \in [0,2\pi], \phi_2 \in [0,2\pi]$ are
coordinates on $S^3$ \cite{s31}, \cite{s32}, \cite{s33}, and $\eta
\in [0,\pi]$ gives the extension to $S^4$. The radius of the base sphere
is $R_0$. In this subsection we propose a field theoretical model
for which such topologically nontrivial solitons may be found in an
exact form. Let us notice that topological defects of this kind are
relevant for $SU(2)$ Yang-Mills theory in its euclideanized $3+1$
dimensional version.
\\
The particular form of the Lagrangian density is
\begin{equation}
L=\left(\frac{4\xi}{(1+\xi^2)^2(1+|u|^2)^2} \right)^{\frac{7}{5}}
H_{(3)}^{\frac{7}{10}}. \label{model fin}
\end{equation}
The value of the $q$ parameter has been chosen to render the
energies of the
solutions finite. For the base space
 $S^4\times \RR$ such Lagrangians are admissible. This is
due to the fact that the radius of the sphere fixes the scale in the
model. Different values of the radius correspond to different
theories. On the other hand in $\RR^4$ base space such Lagrange
densities give non-scale invariant static energies and soliton
solutions would be unstable according to the Derick argument.
From now on we
neglect $R_0$ in our calculations as one may always recover it
using the dimensional analysis. Moreover, we assume the
following Ansatz for static solutions
\begin{equation}
u=f(z) e^{i(n\phi_1+m\phi_2)}, \;\;\;\; \xi=\xi (\eta).
\end{equation}
Obviously, the complex field $u$ is just a Hopf map with $Q_H=\pm
mn$ topological index whereas the scalar $\xi$ provides an extension
to $S^4$. Then the gradients are
\begin{equation}
\nabla u = \frac{1}{\sin \eta} \left[2\sqrt{z(1-z)} f_z,
\frac{inf}{\sqrt{1-z}}, \frac{imf}{\sqrt{z}},0\right], \;\;\;\;
\nabla \xi = [0,0,0,\xi_{\eta}].
\end{equation}
Notice that $\nabla u \nabla \xi= \nabla \bar{u} \nabla \xi \equiv
0$. Hence, $$ H_{(3)}= (\nabla \xi)^2 \left( (\nabla u \nabla
\bar{u} )^2 - (\nabla u)^2 (\nabla \bar{u})^2\right)
$$ and
\begin{equation}
H_{(3)}=16 \frac{z(1-z)}{\sin^4 \eta}
f^2f^2_z\xi^2_{\eta}\left(\frac{n^2}{1-z} +\frac{m^2}{z} \right).
\label{H3}
\end{equation}
Thus,
\begin{equation}
\vec{h}=\frac{8\sqrt{z(1-z)} \xi^2_{\eta} f_zf}{\sin^3 \eta} \left(
\begin{array}{c}
f\left(\frac{n^2}{1-z} +\frac{m^2}{z} \right)\\
-2in\sqrt{z}f_z \\
-2im\sqrt{1-z}f_z \\
0
\end{array} \right) e^{-i(n\phi_1+m\phi_2)}
\end{equation}
and
\begin{equation}
\vec{k}=\frac{32 z(1-z) f_z^2f^2}{\sin^4 \eta}
\left(\frac{n^2}{1-z} +\frac{m^2}{z} \right) \left(
\begin{array}{c}
0\\
0\\
0\\
\xi_{\eta}
\end{array} \right) .
\end{equation}
The static equations of motion are
\begin{equation}
\nabla \vec{\mathcal{H}}=0, \;\;\; \nabla \vec{\mathcal{K}}=0.
\end{equation}
$$
\hspace*{-2cm} \vec{\mathcal{H}}=
\left(\frac{4\xi}{(1+\xi^2)^2(1+f^2)^2} \right)^{\frac{2}{5}}
\frac{8\sqrt{z(1-z)} \xi^2_{\eta} f_zf}{\sin^3 \eta} \left[
\frac{16 \xi_{\eta}^2}{\sin^4 \eta} z(1-z)f^2f^2_z
\left(\frac{n^2}{1-z} +\frac{m^2}{z} \right) \right]^{-3/10}
\times$$
\begin{equation}
\hspace*{9cm} \left(
\begin{array}{c}
f\left(\frac{n^2}{1-z} +\frac{m^2}{z} \right)\\
-2in\sqrt{z}f_z \\
-2im\sqrt{1-z}f_z \\
0
\end{array} \right) e^{-i(n\phi_1+m\phi_2)}
\end{equation}
$$
\hspace*{0cm}
\vec{\mathcal{K}}=\left(\frac{4\xi}{(1+\xi^2)^2(1+f^2)^2}
\right)^{\frac{2}{5}} \frac{32 z(1-z) f_z^2f^2}{\sin^4 \eta}
\left(\frac{n^2}{1-z} +\frac{m^2}{z} \right) \left[ \frac{16
\xi_{\eta}^2}{\sin^4 \eta} z(1-z)f^2f^2_z \left(\frac{n^2}{1-z}
+\frac{m^2}{z} \right) \right]^{-3/10} \times $$
\begin{equation}
\hspace*{9cm} \left(
\begin{array}{c}
0\\
0\\
0\\
1
\end{array} \right).
\end{equation}
Then,  we derive an ordinary differential equation for $f$
$$
\partial_z\left[ z(1-z)
\left(\frac{4\xi}{(1+\xi^2)^2(1+f^2)^2} \right)^{\frac{2}{5}}
  f^2f_z \left[ \frac{16
\xi_{\eta}^2}{\sin^4 \eta} z(1-z)f^2f^2_z \left(\frac{n^2}{1-z}
+\frac{m^2}{z} \right) \right]^{-3/10} \left(\frac{n^2}{1-z}
+\frac{m^2}{z} \right) \right]- $$
\begin{equation}
z(1-z) \left(\frac{4\xi}{(1+\xi^2)^2(1+f^2)^2}
\right)^{\frac{2}{5}}
 ff_z^2 \left[ \frac{16
\xi_{\eta}^2}{\sin^4 \eta} z(1-z)f^2f^2_z \left(\frac{n^2}{1-z}
+\frac{m^2}{z} \right) \right]^{-3/10} \left(\frac{n^2}{1-z}
+\frac{m^2}{z} \right)=0
\end{equation}
and for $\xi$
\begin{equation}
\partial_{\eta} \left( \left(\frac{4\xi}{(1+\xi^2)^2(1+f^2)^2}
\right)^{\frac{2}{5}} \frac{ f_z^2f^2}{\sin \eta} \left(n^2z
+m^2(1-z) \right) \left[ \frac{16 \xi_{\eta}^2}{\sin^4 \eta}
z(1-z)f^2f^2_z \left(\frac{n^2}{1-z} +\frac{m^2}{z} \right)
\right]^{-3/10} \right)=0.
\end{equation}
They may be integrated to the following first order equations
\begin{equation}
\frac{2}{(1+f^2)^2} ff_z=\frac{C_1}{(n^2z+m^2(1-z))^{\frac{7}{4}}}
\end{equation}
\begin{equation}
\frac{2}{(1+\xi^2)^2} \xi \xi_{\eta}= \frac{C_2}{\sqrt{\sin
\eta}}.
\end{equation}
Here $C_1, C_2$ are integration constants. The topologically
nontrivial solutions are
\begin{equation}
\frac{1}{1+f^2}=\frac{(nm)^{\frac{3}{2}}}{n^{\frac{3}{2}}-m^{\frac{3}{2}}}
\left[ \frac{1}{(m^2(1-z)+n^2z)^{\frac{3}{4}}} -
\frac{1}{n^{\frac{3}{2}}} \right],
\end{equation}
\begin{equation}
\frac{1}{1+\xi^2}=1-\frac{\sqrt{2\pi}}{\Gamma^2[\frac{1}{4}]}
\int_0^{\eta} \frac{d\eta'}{\sqrt{\sin \eta'}}.
\end{equation}
The last integral can be easily expressed via the elliptic
function of the first type. The boundary conditions have been
chosen as
\begin{equation}
f(z=1)=\infty, \;\; f(z=0)=0, \;\;\; \mbox{and} \;\;\; \xi
(\eta=\pi)=\infty, \;\;  \xi (\eta=0)=0.
\end{equation}
The total energy reads (if $m^2 \neq n^2$)
\begin{equation}
E=2\pi^2 4^{7/5} \left( \frac{3 \sqrt{2\pi}}{4
\Gamma^2[\frac{1}{4}]}\right)^{2/5} (nm)^{3/5} \left(
\frac{n^2-m^2}{n^{3/2}-m^{3/2}} \right)^{2/5}.
\end{equation}
The complicated dependence on the Hopf charge carried by the
original Hopf map is an artefact of the particular value of the
power in the Lagrangian.

\vspace*{0.2cm}

\noindent

Using the suspended Hopf maps $S^4 \rightarrow S^3$ obtained above, we
are able to construct configurations with nontrivial $\pi_5(S^4)$.
The procedure is analogous, we suspend the suspended Hopf maps
deriving configurations we call 2-suspended Hopf maps. Namely,
\begin{equation}
\vec{M}: \left(
\begin{array}{c}
R_0\sqrt{z} \cos \phi_2 \sin \eta \sin \theta\\
R_0\sqrt{z} \sin \phi_2 \sin \eta \sin \theta\\
R_0\sqrt{1-z} \cos \phi_1 \sin \eta \sin \theta\\
R_0\sqrt{1-z} \sin \phi_1 \sin \eta \sin \theta\\
R_0\cos \eta \sin \theta \\
R_0\cos \theta
\end{array} \right) \longrightarrow \left(
\begin{array}{c}
n_1 \sin \eta \sin \theta\\
n_2 \sin \eta \sin \theta\\
n_3 \sin \eta \sin \theta\\
\cos \eta \sin \theta\\
\cos \theta
\end{array} \right),
\end{equation}
where $\theta \in [0,\pi]$ is a new angle on $S^5$. In this case
one should investigate the following Lagrange density
\begin{equation}
L=\omega(u,\bar{u},v,\bar{v}) H_{(4)}^{q}, \label{model n4 sol}
\end{equation}
where $H_{(4)}$ is a quantity obtained from the pullback of the
volume 4-form on $S^4$. Of course, one may proceed further and
consider higher suspended Hopf maps.
\subsection{$\pi_{2n}(S^{n})$, $n\geq 2$ textures}

In the case when the spatial dimension of the base space-time is
twice as big as the dimension of the target space $d=2n$,
the Hodge dual tensor has the same rank as the original
$h_{\mu_1...\mu_n}$ tensor
\begin{equation}
^*h_{\mu_1...\mu_n}=\frac{1}{n!}\epsilon_{\mu_1...\mu_{2n}}
h^{\mu_{n+1}...\mu_{2n}}.
\end{equation}
Moreover, the scaling invariant action is quadratic in $H_{(n)}$,
giving equations of motion linear in the canonical momenta.
\\
It is interesting to observate that energetically
nontrivial solutions cannot be of the self-dual type. The proof is
as follows. Let us assume that soliton solutions solve the self-dual
equation
\begin{equation}
h_{\mu_1...\mu_n}= \pm \, ^* h_{\mu_1...\mu_n}.
\end{equation}
On the other hand, the static energy reads
\begin{equation}
E=\int d^{2n}x h_{\mu_1...\mu_n}^2=\frac{1}{2}\int d^{2n}x \left(
h_{\mu_1...\mu_n} \pm \, ^*h_{\mu_1...\mu_n}\right)^2 \mp \int
d^{2n}x h_{\mu_1...\mu_n} \,^*h^{\mu_1...\mu_n}.
\end{equation}
For self-dual configurations the first integral vanishes, and
\begin{equation}
E= \frac{1}{n!} \int d^{2n}x \epsilon_{\mu_1...\mu_{2n}}
h_{\mu_1...\mu_n} h^{\mu_{n+1}...\mu_{2n}}.
\end{equation}
However, this expression is identically zero for our $h_{\mu_1...\mu_n}$ tensor.
Therefore all self-dual solutions are in the vacuum.
\\
Thus, it follows that nontrivial solitons of $\pi_4(S^2) \cong
\ZZ_2$ in the AFZ model, which are relevant for the maximal
Abelian projection of the $SU(2)$ Yang-Mills theory on the
Euclidean space $\RR^4$, cannot be of the self-dual nature.
\\
The same happens for the three dimensional integrable models on
$(6+1)$ Minkowski space-time. This last class of theories is very
special, as static solutions may be divided into disjoint classes
as they are maps from $S^6$ onto $S^3$ and classified by means of
a rather exotic topological index $Q \in \pi_6(S^3) \cong
\ZZ_{12}$. Using the fact proved in this
subsection we conclude that such defects are not of the Bogomolny
type, at least as long as one restricts to the integrable pullback
models.
\section{Nontrivial $\pi_{n}(S^{n})$, $n\geq 3$ textures}
\subsection{$\pi_{4}(S^{4})$ textures}
There is a way how to construct nontrivial (non purely
topological) models, using the scalars $H_{n}$ defined in our
procedure, which possess solitons carrying the generalized baryon
number $Q \in \pi_n (S^n)$. These textures are not exotic and have
been studied before \cite{radu1}, \cite{radu2}. However, we
propose new models for such a class of defects which,
additionally, appear to be of the Bogomolny type satisfying some
first order Bogomolny equations. Moreover, they will be given in
an exact form. Such topological defects can be viewed as higher
dimensional (in the base space as well as the target space)
generalizations of the dilaton-Yang-Mills system with topological
index taking values in the $\pi_3(S^3)$ homotopy group
\cite{dilaton}, \cite{bizon}.
\\
As a first example, we consider a dilaton extension of the three
dimensional pullback model in $(4+1)$ dimensional space-time
\begin{equation}
\mathcal{L}=\int d^4 x \left( \frac{1}{2} \phi^2_{\mu}+e^{2\phi}
\omega^2(u\bar{u}) \sigma^2 (\xi) h_{\mu \nu \rho}^2 \right),
\label{soliton 4d mod}
\end{equation}
where $\phi$ is the dilaton field while $\omega$ and $\sigma$ are
arbitrary scalar functions depending on the modulus of the complex
and scalar field, respectively. For convenience, we slightly
changed the definition of $h_{\mu \nu \rho}$ multiplying it by $i$,
which makes the tensor real. The static Bogomolny equation can be
found via the standard trick
\begin{equation} \label{E-dil}
E=\int d^4 x \left( \frac{1}{\sqrt{2}} \phi_{i} - e^{\phi} \omega
(u\bar{u}) \sigma (\xi) \,^* h_{i} \right)^2 + \sqrt{2} \sqrt{3!} \int d^4 x
e^{\phi} \omega (u\bar{u}) \sigma (\xi)\;^* h_{i} \phi^{i},
\end{equation}
where
\begin{equation}
^*h_{i}=\frac{1}{3!} \epsilon_{ijkl} h^{jkl}
\end{equation}
is the pertinent Hodge dual. The first term at the r.h.s. of Eq. (\ref{E-dil})
just provides the Bogomolny equation
\begin{equation}
\frac{1}{\sqrt{2}} \phi_{i} - \frac{e^{\phi}\omega (u\bar{u})
\sigma (\xi)}{\sqrt{3!}} \epsilon_{ijkl} h^{jkl}=0. \label{4d bog}
\end{equation}
Then,
\begin{equation}
E \geq \frac{1}{\sqrt{3}} \int d^4 x e^{\phi} \omega (u\bar{u})
\sigma (\xi)\epsilon_{ijkl} \phi^{i} h^{jkl} = \frac{1}{\sqrt{3}}
\int d^4 x \nabla^i \left( e^{\phi} \omega (u\bar{u}) \sigma (\xi)
\epsilon_{ijkl} h^{jkl} \right)
\end{equation}
\begin{equation}
\hspace*{6.2cm} = \frac{1}{\sqrt{3}} \oint_{\Sigma} ds_{3}^{i}
e^{\phi} \omega (u\bar{u}) \sigma (\xi) \epsilon_{ijkl} h^{jkl}.
\end{equation}
Solitons which we expect to live in this system are stable under
scale transformations. This is due to the fact that the gradient
terms for the $\phi$ and $h_{\mu \nu \rho}$ fields in the static energy
scale oppositely
\begin{equation} E[\lambda]=\lambda^2
E_1+\frac{1}{\lambda^2}E_2.
\end{equation}
As the target space of the model is $4$ dimensional, we get
objects with can be associated with nontrivial representatives of
the $\pi_4(S^4)$ homotopy class.
\\
Let us now prove that there are exact soliton configurations which
saturate the Bogomolny bound derived above. The full equations of
motion for the system (\ref{soliton 4d mod}) are
\begin{equation}
\partial_{\mu}^2 \phi - 2 e^{2\phi} \omega^2 \sigma^2 h_{\mu \nu
\rho}^2=0
\end{equation}
\begin{equation}
\partial_{\mu} \left( e^{2\phi} \omega h^{\mu} \right)=0
\end{equation}
\begin{equation}
\partial_{\mu} \left( e^{2\phi} \sigma k^{\mu} \right)=0,
\end{equation}
with $h^{\mu}, k^{\mu}$ given by (\ref{vec h n3}), (\ref{vec k n3}).
Their static version reads
\begin{equation}
\nabla^2 \phi - 2 \cdot 3! e^{2\phi} \omega^2\sigma^2 (\nabla
\xi)^2 \left[(\nabla u \nabla \bar{u})^2 - (\nabla u)^2 (\nabla
\bar{u})^2\right]=0 \label{4dil static 1}
\end{equation}
\begin{equation}
\nabla \left( \omega \vec{h} \right)=0, \;\;\;\;\;\;\;\;\;
\vec{h}=2 (\nabla \xi)^2 \left[(\nabla u \nabla \bar{u}) \nabla
\bar{u} - (\nabla \bar{u})^2 \nabla u \right] \label{4dil static
2}
\end{equation}
\begin{equation}
\nabla \left( \sigma \vec{k} \right)=0, \;\;\;\;\;\;\;\;\; \vec{k}=2
 \left[(\nabla u \nabla \bar{u})^2 - (\nabla u)^2 (\nabla
\bar{u})^2 \right] \nabla \xi, \label{4dil static 3}
\end{equation}
where we assumed the following Ansatz
\begin{equation}
\phi=\phi (r), \;\;\;\; u=u(\theta, \varphi)=f(\theta)
e^{im\varphi}, \;\;\;\; \xi=\xi (\alpha),
\end{equation}
where $r, \alpha, \theta, \varphi$ are spherical coordinates in 4
dimensions such that
$$
\nabla = \left[\partial_r, \frac{1}{r}\partial_{\alpha},
\frac{1}{r\sin \alpha} \partial_{\theta}, \frac{1}{r\sin \alpha
\sin \theta}
\partial_{\varphi} \right]
$$
Hence,
\begin{equation}
\nabla u = \frac{1}{r\sin \alpha} \left[0,0,f_{\theta},
\frac{imf}{\sin \theta} \right] e^{im\varphi}, \;\;\;\;\; \nabla \xi
= \frac{1}{r} \left[0,\xi_{\alpha},0,0\right]
\end{equation}
and
\begin{equation}
\vec{h}=\frac{4\xi^2_{\alpha} m ff_{\theta}}{r^5\sin^3 \alpha \sin
\theta} e^{-im\varphi} \left[0,0, \frac{mf}{\sin
\theta},-if_{\theta} \right]
\end{equation}
\begin{equation}
\vec{k}=\frac{8m^2f^2f^2_{\theta}}{r^5\sin^4 \alpha \sin^2
\theta}[0,\xi_{\alpha},0,0].
\end{equation}
The homogenous equations (\ref{4dil static 2}), (\ref{4dil static 3})
lead to
\begin{equation}
\partial_{\theta} \left( \ln \frac{\omega ff_{\theta}}{\sin \theta}
\right)=0, \;\;\;\;\;\; \partial_{\alpha} \left( \frac{\sigma
\xi_{\alpha}}{\sin^2 \alpha} \right)=0
\end{equation}
or
\begin{equation}
\frac{\omega ff_{\theta}}{\sin \theta}=c_1, \;\;\;\;\;\;
\frac{\sigma \xi_{\alpha}}{\sin^2 \alpha}=c_2,
\end{equation}
where $c_1,c_2$ are integration constants. Substituting these
formulas into the equation for the dilaton (\ref{4dil static 1}) we get
\begin{equation}
\frac{1}{r^3} \partial_r (r^3\phi_r) -3\cdot 4^2m^2
\frac{e^{2\phi}}{r^6} \left( \frac{\omega ff_{\theta}}{\sin
\theta}\right)^2 \left( \frac{\sigma \xi_{\alpha}}{\sin^2 \alpha}
\right)^2=0
\end{equation}
\begin{equation}
r^3 \partial_r (r^3\phi_r) -3\cdot 4^2m^2 e^{2\phi}c_1^2c_2^2=0.
\end{equation}
Then
\begin{equation}
\phi_{xx}-12m^2c_1^2c_2^2e^{2\phi}=0, \;\;\;\;\;\;\;\;
x=\frac{1}{r^2},
\end{equation}
and after integration
\begin{equation}
\frac{1}{2} \phi_x^2=c_3+6 m^2c_1^2c_2^2e^{2\phi}.
\end{equation}
For finite energy solutions one has to put $c_3=0$.  In addition,
assuming topologically nontrivial boundary conditions $\phi (x=0)=0$
and $\phi (x=\infty)=-\infty$, we obtain the solution
\begin{equation}
\phi=-\ln \left[ \sqrt{12}mc_1c_2 \left( \frac{1}{r^2} +
\frac{1}{m\sqrt{12}c_1c_2} \right)\right].
\end{equation}
Specific expressions for $u$ and $\xi$ can be found if we
specify the coupling functions $\omega$ and $\sigma$. For example, we
consider
\begin{equation}
\omega=\frac{1}{(1+|u|^2)^2}, \;\;\;\; \sigma=e^{\kappa \xi}.
\end{equation}
Again, in order to get topologically nontrivial configurations, $u$
should be a map from $S^2$ to $S^2$. This requires that the modulus
covers the positive real half-axis $f(\theta =0)=0, f(\theta
=\pi)=\infty$. Then we obtain that $c_1=\frac{1}{4}$ and
\begin{equation}
f=\tan \frac{\theta}{2} \;\;\;\;\;\;\Rightarrow \;\;\;\;\;\;
u(\theta,\varphi)= \tan \frac{\theta}{2} e^{im\varphi}
\end{equation}
i.e., a map $S^2 \rightarrow S^2$ with winding number $m$.
Finally,
\begin{equation}
\frac{1}{\kappa} e^{\kappa \xi}=\frac{c_2}{2}\left( \alpha -
\frac{1}{2}\sin 2\alpha \right) +d_2.
\end{equation}
Now, the boundary conditions are $\xi(\alpha=0)=-\infty$ and
$\xi(\alpha=\pi)=0$. Therefore, $d_2=0$, $c_2=\frac{2}{\pi \kappa}$
and
\begin{equation}
\xi = \frac{1}{\kappa} \ln \left[ \frac{1}{\pi} \left(
\alpha-\frac{1}{2} \sin 2\alpha \right) \right].
\end{equation}
Thus, taking into account the values of $c_1, c_2$ we arrive at the
final formula for the dilaton field
\begin{equation}
\phi=-\ln \left[ \frac{\sqrt{3}m}{\pi \kappa} \left( \frac{1}{r^2}
+ \frac{\pi \kappa}{\sqrt{3}m} \right)\right].
\end{equation}
The corresponding energy is
\begin{equation}
E=4\sqrt{3} \frac{m}{\pi \kappa} S_{(3)}=8\sqrt{3} \frac{\pi}{
\kappa} m,
\end{equation}
where $S_{(3)}=2\pi^2$ is area of the three dimensional sphere
with unit radius.
\\
Dilaton extension of the rank three pullback model (\ref{soliton
4d mod}) reveals an interesting similarity to the standard Abelian
gauge system coupled with a dilaton field. In order to see it, let
us consider the following Lagrange density
\begin{equation}
L=
\frac{1}{2} \tilde{\phi}^2_{\mu}-e^{2\tilde{\phi}} F_{\mu \nu}^2.
\label{dual 4d mod}
\end{equation}
where $F_{\alpha \beta}$ is the Abelian gauge field in (4+1)
dimensions. The static, pure electric field equations are
\begin{equation}
\nabla^2 \tilde{\phi} - 4 e^{-2\tilde{\phi}} \vec{E}^2=0
\end{equation}
\begin{equation}
\nabla \cdot \left( e^{-2 \tilde{\phi} } \vec{E} \right) = 2\pi^2
q \delta (r),
\end{equation}
where we use the standard definition $E^i=-F^{0i}$, $i=1,2,3,4$
and assumed vanishing magnetic field. Here $q$ is an electric
charge. The obvious solution is
\begin{equation}
\vec{E}=\frac{q }{r^3} e^{-2 \tilde{\phi}} \hat{r},
\end{equation}
\begin{equation}
\tilde{\phi}=\ln \left[ 2q \left( \frac{1}{r^2} + \frac{1
 }{2q} \right) \right].
\end{equation}
Therefore, the solitons found for model (\ref{soliton 4d mod}) can
be viewed as dual configurations to electric dilaton solutions
generated by a point charge in four space dimensions. Indeed,
there is a dual transformation connecting solutions of the
models (\ref{soliton 4d mod}), (\ref{dual 4d mod})
\begin{equation}
\frac{1}{\sqrt{3}} \omega (u\bar{u}) \sigma (\xi) h_{\mu \nu \rho}  =
\epsilon_{\mu \nu \rho \alpha \beta} e^{2\tilde{\phi}} F^{\alpha
\beta} \;\;\;\; \mbox{and} \;\;\;\; \phi = - \tilde{\phi}.
\label{dual trans}
\end{equation}
The relation between the electric charge and topological index
$m$, which makes this dual transformation correct, is
\begin{equation}
q=\frac{\sqrt{3}}{2} \frac{m}{\pi \kappa}.
\end{equation}
Notice, that the transformation is realized at the Lagrangians
and well as at the solution level. It may be verify using the relation
\begin{equation}
-\frac{1}{2} \left( \epsilon_{\mu \nu \rho \alpha \beta} F^{\alpha \beta} \right)^2= \frac{1}{3!}F_{\alpha \beta}^2,
\end{equation}
where minus sign comes from the Minkowski signature.
Thus, it makes this tranformation a real duality tranformation.

\vspace*{0.2cm}

\noindent {\bf Remark:}
\\
Observe, that a Bogomolny equation similar to (\ref{4d bog}) has
been analyzed for a generalized Goldstone model in (4+1)
dimensions \cite{radu2}, derived as the gauge decoupled limit of
the $SO(4)$ Higgs-Yang-Mills model descended from the $3$-th
member of the Yang-Mills hierarchy on $\RR_4 \times S^8$
\cite{tch1}, \cite{tch2}. However, our model (\ref{soliton 4d
mod}) differs qualitatively as well as quantitatively from that
Goldstone model. First of all the topology of solutions is
different. Here, we are looking for $\pi_4(S^4)$ solitons whereas
Radu-Tchrakian solutions are of $\pi_3(S^3)$ type as the pertinent
4 dimensional isovector $\vec{\phi}$ is assumed to tend to a
vacuum value at the spatial infinity.\footnote{Such a vector maybe
identify with the field content of the rank-3 dilaton model as
$\vec{\phi}=(\phi, \mbox{Re} \, u, \mbox{Im} \, u, \xi)$.}
Moreover, solitons in this model do not saturate the Bogomolny
bound. On the contrary they obey full second order equations of
motion. Additionally, they are not known in exact forms. On the
other hand, numerical solutions presented in \cite{radu2} describe
configurations with more complicated and interesting geometry than
only spherical or axial one. It would be interesting to check
whether such multi-solitons may be found in an exact form in
dilaton rank-3 model (\ref{soliton 4d mod}). One could also
analyze the Goldstone model in (4+1) dimensions from the
generalized integrability point of view. Perhaps, one could
connect the appearance of various soliton solutions with the
existence of families of the infinitely many conserved currents.
\subsection{$\pi_{5}(S^{5})$ textures}

In the same fashion one is able to find a Lagrangian for defects
of the $\pi_5(S^5)$ homotopy type. The Lagrangian relevant in this
context reads
\begin{equation}
\mathcal{L}=\int d^5 x \;\left( \frac{1}{2}\phi_{\mu}^2 - e^{2\phi}
\omega^2 (u\bar{u}) \sigma^2 (v,\bar{v}) \; h_{\mu \nu \rho
\sigma}^2 \right), \label{soliton 5d mod}
\end{equation}
where we use the pullback tensor of $4$-dimensional target space.
The corresponding Bogomolny equations are
\begin{equation}
\frac{1}{\sqrt{2}} \phi_{i} -  \frac{e^{\phi} \omega (u\bar{u})
\sigma (v,\bar{v})}{\sqrt{4!}} \epsilon_{ijklm} h^{jklm}=0,
\end{equation}
and the energy is
\begin{equation}
E=\int d^5 x \left( \frac{1}{\sqrt{2}} \phi_{i} - \sqrt{4!}e^{\phi} \omega
(u\bar{u}) \sigma (v,\bar{v}) \,^* h_{i} \right)^2 + \sqrt{2} \sqrt{4!}\int
d^5 x e^{\phi} \omega (u\bar{u}) \sigma (v\bar{v})\;^* h_{i}
\phi^{i},
\end{equation}
where
\begin{equation}
^*h_{i}=\frac{1}{4!} \epsilon_{ijklm} h^{jklm}.
\end{equation}
Using the Bogomolny equation we find
\begin{equation}
E \geq \frac{1}{2\sqrt{3}} \int d^5 x e^{\phi} \omega (u\bar{u})
\sigma (v,\bar{v})\epsilon_{ijklm} \phi^{i} h^{jklm} =
\frac{1}{2\sqrt{3}} \int d^5 x \nabla^i \left( e^{\phi} \omega
(u\bar{u}) \sigma (v,\bar{v}) \epsilon_{ijklm} h^{jklm} \right)
\end{equation}
\begin{equation}
\hspace*{6.6cm} =\frac{1}{2\sqrt{3}}\oint_{\Sigma} ds_{4}^{i}
e^{\phi} \omega (u\bar{u}) \sigma (v,\bar{v}) \epsilon_{ijklm}
h^{jklm}.
\end{equation}
The energy scales as
\begin{equation}
E[\lambda]=\lambda^3 E_1+\frac{1}{\lambda^3} E_2.
\end{equation}
Static solitons saturating the Bogomolny bound may be obtained in
the same way as before. The equations of motion for our system are
\begin{equation}
\partial_{\mu}^2\phi+2e^{2\phi}\omega^2\sigma^2 h_{\mu \nu \rho
\sigma}^2=0,
\end{equation}
\begin{equation}
\partial_{\mu} \left( e^{2\phi} \omega h^{\mu} \right)=0, \;\;\;\;
\partial_{\mu} \left( e^{2\phi} \sigma k^{\mu} \right)=0,
\end{equation}
and the complex conjugates. Here $h_{\mu}$ and $ k_{\mu}$ are the canonical
momenta conjugate to the $u$ and $v$ fields, respectively. We used that
$u_{\mu}k^{\mu}=\bar{u}_{\mu}k^{\mu}=0$ and
$v_{\mu}h^{\mu}=\bar{v}_{\mu}h^{\mu}=0$. For static configurations we
assume the following Ansatz
\begin{equation}
\phi = \phi (r), \;\;\; u=u(\theta, \varphi), \;\;\; v=v(\beta,
\alpha)
\end{equation}
where $r,\beta, \alpha, \theta, \varphi$ are $5$ dimensional
spherical coordinates giving $$ \nabla = \left[\partial_r,
\frac{1}{r}\partial_{\beta}, \frac{1}{r \sin
\beta}\partial_{\alpha}, \frac{1}{r\sin \beta \sin \alpha}
\partial_{\theta}, \frac{1}{r\sin
\beta \sin \alpha \sin \theta} \partial_{\varphi} \right].$$ Then,
\begin{equation}
-\nabla_r^2 \phi+4! 2e^{2\phi}\omega^2 \sigma^2 \left[ (\nabla
u)^2(\nabla \bar{u})^2 - (\nabla u \nabla \bar{u})^2 \right] \left[
(\nabla v)^2(\nabla \bar{v})^2 - (\nabla v \nabla \bar{v})^2
\right]=0, \label{5d eom1}
\end{equation}
\begin{equation}
\nabla \cdot (\omega \vec{h} )=0, \;\;\;\; \vec{h}= \left[ (\nabla
v)^2(\nabla \bar{v})^2 - (\nabla v \nabla \bar{v})^2 \right] \left[
(\nabla \bar{u})^2 \nabla u - (\nabla u \nabla \bar{u}) \nabla
\bar{u} \right], \label{5d eom2}
\end{equation}
\begin{equation}
\nabla \cdot (\sigma \vec{k} )=0, \;\;\;\; \vec{k}= \left[ (\nabla
u)^2(\nabla \bar{u})^2 - (\nabla u \nabla \bar{u})^2 \right]
\left[ (\nabla \bar{v})^2 \nabla v - (\nabla v \nabla \bar{v})
\nabla \bar{v} \right] \label{5d eom3}
\end{equation}
as $\nabla u \nabla v = \nabla \bar{u} \nabla v = \nabla u \nabla
\bar{v} = \nabla \bar{u} \nabla \bar{v} =0$, $\nabla \phi \nabla
u= \nabla \phi \nabla \bar{u} = \nabla \phi \nabla v = \nabla \phi
\nabla \bar{v}=0$. Equation (\ref{5d eom2}) is easily solved by
the configuration $u=f(\theta)e^{im\varphi}$
\begin{equation}
\frac{\omega ff_{\theta}}{\sin \theta}=c_1.
\end{equation}
In order to solve equation (\ref{5d eom3}) we decompose the complex
field $v$ into two real scalars $v=\Sigma+i\Lambda$ such that
\begin{equation}
\Sigma=\Sigma (\beta), \;\;\;\;\;\; \Lambda=\Lambda (\alpha)
\;\;\;\;\;\Rightarrow \;\;\;\;\; \nabla \Sigma \cdot \nabla \Lambda
=0.
\end{equation}
Therefore, we arrive at two equations
\begin{equation}
\nabla \cdot \left( \sigma_1 \left[ (\nabla u)^2(\nabla \bar{u})^2
- (\nabla u \nabla \bar{u})^2 \right] (\nabla \Lambda)^2 \nabla
\Sigma \right)=0,
\end{equation}
\begin{equation}
\nabla \cdot \left( \sigma_2 \left[ (\nabla u)^2(\nabla \bar{u})^2
- (\nabla u \nabla \bar{u})^2 \right] (\nabla \Sigma)^2 \nabla
\Lambda \right)=0,
\end{equation}
where $\sigma (v,\bar{v})=\sigma_1 (\Lambda) \sigma_2(\Sigma) $.
After some calculations we get
\begin{equation}
\partial_{\alpha} \left( \frac{\sigma_1 \Lambda_{\alpha}}{\sin^2 \alpha}
\right)=0, \;\;\; \partial_{\beta} \left( \frac{\sigma_2
\Sigma_{\beta}}{\sin^3 \beta} \right)=0,
\end{equation}
with two obvious solutions
\begin{equation}
\frac{\sigma_1 \Lambda_{\alpha}}{\sin^2 \alpha}=c_2, \;\;\;\;
\frac{\sigma_2 \Sigma_{\beta}}{\sin^3 \beta}=c_3.
\end{equation}
The first order ordinary differential equations are easily solved if
we specify particular forms of the scalar functions $\omega,
\sigma_1, \sigma_2$. For example
\begin{equation}
\omega=\frac{1}{(1+|u|^2)^2}, \;\;\;\; \sigma_1=e^{\kappa \Lambda},
\;\;\;\; \sigma_2=e^{\zeta \Sigma}
\end{equation}
give
\begin{equation}
u(\theta,\varphi)= \tan \frac{\theta}{2} e^{im\varphi},
\end{equation}
\begin{equation}
\Lambda = \frac{1}{\kappa} \ln \left[ \frac{1}{\pi} \left(
\alpha-\frac{1}{2} \sin 2\alpha \right) \right].
\end{equation}
\begin{equation}
\Sigma=\frac{1}{\zeta} \ln \left[ \frac{3}{4} \left( \frac{2}{3}
-\cos \beta +\frac{1}{3} \cos^3 \beta \right) \right],
\end{equation}
with $c_1=1/4, c_2=2/\pi \kappa, c_3=3/4\zeta$ providing a
topologically nontrivial configuration. Now, we return to the last
remaining equation (\ref{5d eom1})
\begin{equation}
\frac{1}{r^4} \partial_r (r^4\phi_r) -4!2e^{2\phi}
\frac{4^2m^2}{r^8}  \left( \frac{\omega ff_{\theta} }{\sin \theta}
\right)^2 \left( \frac{\sigma_1 \Lambda_{\alpha}}{\sin^2 \alpha}
\right)^2 \left( \frac{\sigma_2 \Sigma_{\beta}}{\sin^3 \beta}
\right)^2=0.
\end{equation}
Hence
\begin{equation}
\frac{1}{r^4} \partial_r (r^4\phi_r) -4!2e^{2\phi} \frac{4^2m^2
c_1^2c_2^2c_3^2}{r^8}=0 \;\;\;\;\Rightarrow\;\;\;\; r^4
\partial_r (r^4\phi_r) - \frac{3^32^2m^2}{\pi^2\kappa^2 \zeta^2}e^{2\phi}=0.
\end{equation}
Finally, the dilaton solution is
\begin{equation}
\phi=-\ln \left[ \frac{2\sqrt{3}m}{\pi \kappa \zeta} \left(
\frac{1}{r^3} + \frac{\pi \kappa \zeta}{2\sqrt{3}m} \right)\right]
\end{equation}
and the energy is
\begin{equation}
E=\frac{3^3 \, 2 \, \sqrt{3}}{\pi \kappa \zeta} m
S_{(4)}=\frac{3^24^2 \sqrt{3}\pi}{\kappa \zeta} m,
\end{equation}
where $S_{(4)}= \frac{8}{3}\pi^2$. As we see, such a
$5$-dimensional dilaton solution is a simple extension of the $4$
dimensional case. The additional target space degree of freedom
i.e., the field $\Sigma$, depends on the new coordinates. The
generalization to higher dimensions, that is, to higher homotopy
groups, is obvious.

\vspace*{0.2cm}

\noindent An alternative set-up for solitons of this type is given
by the following Lagrangian containing the dilaton field
\begin{equation}
L=\int d^5 x \; g^2_1(u\bar{u}) h_{\mu \nu}^2 - g^2_2(u\bar{u})
\omega^2(v\bar{v}) \sigma^2(\xi) \; h_{\mu \nu \rho}^2,
\end{equation}
where $h_{\mu \nu}$ depends on $u,\bar{u}$ while $h_{\mu \nu \rho}$
on $v,\bar{v},\xi$ fields respectively. Once again we use the static
energy to derive the static Bogomolny equation
\begin{equation}
E= \int d^5 x \left(g_1(u\bar{u}) h_{ij} -\sqrt{3} g_2(u\bar{u})
\omega(v\bar{v}) \sigma(\xi) \,^* h_{ij} \right)^2 + 2\sqrt{3}\int d^5 x
g_1(u\bar{u}) g_2(u\bar{u})\omega(v\bar{v}) \sigma(\xi) \;^* h_{ij}
h^{ij},
\end{equation}
where
\begin{equation}
^*h_{ij}=\frac{1}{3!} \epsilon_{ijklm} h^{klm}
\end{equation}
is the pertinent Hodge dual of the static part of the $h_{\mu \nu \rho}$
tensor. Therefore, the Bogomolny equation reads
\begin{equation}
 g_1(u\bar{u}) h_{ij} -  \frac{1}{2\sqrt{3}}g_1(u\bar{u}) g_2(u\bar{u})\omega(v\bar{v})
\sigma(\xi) \epsilon_{ijklm} h^{klm}=0
\end{equation}
and
\begin{equation}
E \geq \frac{2\sqrt{3}}{3!} \int d^5 x g_1(u\bar{u}) g_2(u\bar{u})
\omega(v\bar{v}) \sigma(\xi) \epsilon_{ijklm} h^{ij} h^{klm} =
\frac{1}{\sqrt{3}} \int d^5 x \nabla^i \left( \tilde{g} (u\bar{u})u \;
\omega(v\bar{v}) \sigma(\xi)\; \epsilon_{ijklm} \bar{u}^{j} h^{klm}
\right)=
\end{equation}
\begin{equation}
\hspace*{5cm} \frac{1}{\sqrt{3}} \oint_{\Sigma} ds^{i}_{(4)} \tilde{g}
(u\bar{u})u \;  \omega(v\bar{v}) \sigma(\xi)\; \epsilon_{ijklm}
\bar{u}^{j} h^{klm}, \;\;\;\; \mbox{where} \;\;\;\;
\tilde{g}'u\bar{u}+\tilde{g}=g_2
\end{equation}
and the prime denotes differentiation with respect to $u\bar{u}$.
Also in this model we avoid the scaling instabilities, because the static
energy scales as
\begin{equation}
E[\lambda]=\lambda E_1+\frac{1}{\lambda} E_2.
\end{equation}
Of course, using other pullback tensors we can derive theories with
arbitrary $\pi_n(S^n)$ solitons.
\\
It is worth stressing that these models provide standard second
order dynamical equations of motion. This is in contrast to
models with exotic textures, for which the Derrick theorem enforces
nonlinear (in the canonical momenta) field equations. As a
consequence, one gets equations which in some regions are not
hyperbolical, leading to serious problems as far as time evolution of
solitons is concerned.
\section{Conclusions}
The present paper further develops and applies
the Generalized Zero Curvature approach of Ref. \cite{joaquin1}.
Its main results are the construction of several families
of integrable models living on $(d+1)$ dimensional space-times and
characterized by arbitrary target space manifolds of dimension
$n$, as well as the detailed investigation of their solutions.
Such models are integrable in the sense that they possess the
generalized zero curvature formulation, infinitely many conservation
laws and, in those particular cases we have checked, admit
infinitely many static solutions with a nontrivial topological
structure.
The origin of the conserved currents has also been clarified. They
are generated by the volume preserving diffeomorphisms on the target
space.
\\
All these models are built from the pullback of the pertinent
target space volume $n$-form into the $(d+1)$ (Minkowski) base space,
and may be
multiplied by a density function depending entirely on the fields and
not on their derivatives. Therefore the models are trivial if $d
\leq n$. On the other hand, if $d
> n$, the integrable models describe nontrivial highly nonlinear
field theories. Therefore, the solitons found in these models are,
in general, of a nonstandard (exotic) type from the topology point of
view. In fact, hopfions as well as suspended Hopf maps (higher
suspended Hopf maps as well) can be derived as finite energy
solutions to those models. These configurations are probably not of
the Bogomolny type and are not given by any first order field
equations. The first argument is based on the observation that the
dependence of the energy on the topological charge is non-linear.
The second argument is given in this work and shows that, at least
for some of the models, self-dual configurations contribute to the
vacuum manifold. However, this problem is by no means solved and
definitely requires further studies. On the other hand, the fact
that the higher dimensional nonlinear models derived here may be
solved exactly, in spite of the lack of Bogomolny equations, is very
intriguing and indicates that the generalized integrability can
serve, to some extent, as an alternative to the Bogomolny limit.
\\
Using the pullback tensors, we have been able to construct models
with exact solitons with more standard topological content.
Infinitely many finite energy static solutions with topological
index belonging to $\pi_4(S^4)$ as well as $\pi_5(S^5)$ have been
obtained. And in contrast to the previous exotic solutions, they
emerge from their pertinent Bogomolny equations.
\\
One possible direction of further investigation consists in the use of
more general ansaetze. Indeed, all solutions presented here have the property
that the gradients of different fields are perpendicular, which simplifies the
calculations significantly. Relaxing this condition may give rise to more
general solutions with a more complicated geometry.
\\
Another interesting question concerns the possibility to embed these
intergrable models into more complicated and, at the same time, more physically
relevant theories, as discussed in Section 5.
This possibility has been studied in detail for the
case of Yang--Mills dilaton theory \cite{we YMdil},
which contains an integrable submodel
defined by abelian projection. This submodel is of the type discussed in
Section 7, but based on the topological index $\pi_3(S^3)$. In that case it
turns out that the soliton solutions are stable in the submodel (the obey
a Bogomolny bound), but unstable, sphaleron-type solutions in the full
Yang--Mills dilaton theory.
Therefore, in the case of the embedding of integrable models into
larger theories the question of stability is rather nontrivial and certainly
requires a detailed study. More generally, possible embeddings into
Yang--Mills type theories and their collective excitations may lead to
applications with some experimental relevance (e.g., to
condensates, or gluonic extended field configurations). These issues will be
further investigated in the future.
\\
In any case, we think that the rather large class of integrable models
together with their explicit solutions presented in this paper are of
interest and will lead to further applications. Let us finally mention that
the conjecture relating intergrability and solvability for higher-dimensional
integrable field theories continues to hold for the class of models studied in
this paper. Indeed, for any theory which possesses infinitely many conserved
quantitites, infinitely many exact solutions may be found by simple
quadratures, once an adequate ansatz has been chosen. This points towards the
existence of a deep relation between integrability and solvability also in
higher dimensions, analogous to the case of 1+1 dimensional field theories.

\section*{Acknowledgements}

C.A., P.K. and J.S.-G. thank MCyT (Spain) and FEDER
(FPA2005-01963), and support from
 Xunta de Galicia (grant PGIDIT06PXIB296182PR and Conselleria de
Educacion). A.W. acknowledges support from the Foundation for
Polish Science FNP (KOLUMB programme 2008/2009) and Ministry of
Science and Higher Education of Poland grant N N202 126735
(2008-2010).

\end{document}